\documentclass[sigconf]{acmart}
\usepackage{amsmath}

\AtBeginDocument{%
  \providecommand\BibTeX{{%
    \normalfont B\kern-0.5em{\scshape i\kern-0.25em b}\kern-0.8em\TeX}}}

\setcopyright{acmlicensed}
\copyrightyear{2021}
\acmYear{2021}
\acmDOI{10.1145/3450613.3456811}

\acmConference[UMAP '21]{Proceedings of the 29th ACM Conference on User Modeling, Adaptation and Personalization}{June 21--25, 2021}{Utrecht, Netherlands}
\acmBooktitle{Proceedings of the 29th ACM Conference on User Modeling, Adaptation and Personalization (UMAP '21), June 21--25, 2021, Utrecht, Netherlands}
\acmPrice{15.00}
\acmISBN{978-1-4503-8366-0/21/06}

\begin{document}

\title[Hassan, et al.]{Learning to Trust: Understanding Editorial Authority and Trust in Recommender Systems for Education}


\author{Taha Hassan}
\affiliation{%
  \institution{Department of Computer Science, Virginia Tech}
  \city{Blacksburg}
  \state{Virginia}}
\email{taha@vt.edu}

\author{Bob Edmison}
\affiliation{%
  \institution{Department of Computer Science, Virginia Tech}
  \city{Blacksburg}
  \state{Virginia}}
\email{kedmison@vt.edu}

\author{Timothy Stelter}
\affiliation{%
  \institution{Department of Computer Science, Virginia Tech}
  \city{Blacksburg}
  \state{Virginia}}
\email{tstelter@vt.edu}

\author{D. Scott McCrickard}
\affiliation{%
  \institution{Department of Computer Science, Virginia Tech}
  \city{Blacksburg}
  \state{Virginia}}
\email{mccricks@vt.edu}

\renewcommand{\shortauthors}{Hassan, et al.}

\begin{abstract}
Trust in a recommendation system (RS) is often algorithmically incorporated using implicit or explicit feedback of user-perceived trustworthy social neighbors, and evaluated using user-reported trustworthiness of recommended items. However, real-life recommendation settings can feature group disparities in trust, power, and prerogatives. Our study examines a complementary view of trust which relies on the editorial power relationships and attitudes of all stakeholders in the RS application domain. We devise a simple, first-principles metric of \textit{editorial authority}, i.e., user preferences for recommendation sourcing, veto power, and incorporating user feedback, such that one RS user group confers trust upon another by ceding or assigning editorial authority. In a mixed-methods study at Virginia Tech, we surveyed faculty, teaching assistants, and students about their preferences of editorial authority, and hypothesis-tested its relationship with trust in algorithms for a hypothetical `Suggested Readings' RS. We discover that higher RS editorial authority assigned to students is linked to the relative trust the course staff allocates to RS algorithm and students. We also observe that course staff favors higher control for the RS algorithm in sourcing and updating the recommendations long-term. Using content analysis, we discuss frequent staff-recommended student editorial roles and highlight their frequent rationales, such as perceived expertise, scaling the learning environment, professional curriculum needs, and learner disengagement. We argue that our analyses highlight critical user preferences to help detect editorial power asymmetry and identify RS use-cases for supporting teaching and research.
\end{abstract}

\ccsdesc[500]{Human-centered computing~HCI theory, concepts and models}
\ccsdesc[300]{Human-centered computing~User models}
\ccsdesc[300]{Human-centered computing~Social recommendation}
\ccsdesc[100]{Information systems~Personalization}

\keywords{recommendation, education, context, trust, interpretation}


\maketitle

\begin{figure*}[t]
  \centering
  \includegraphics[width=0.8\linewidth]{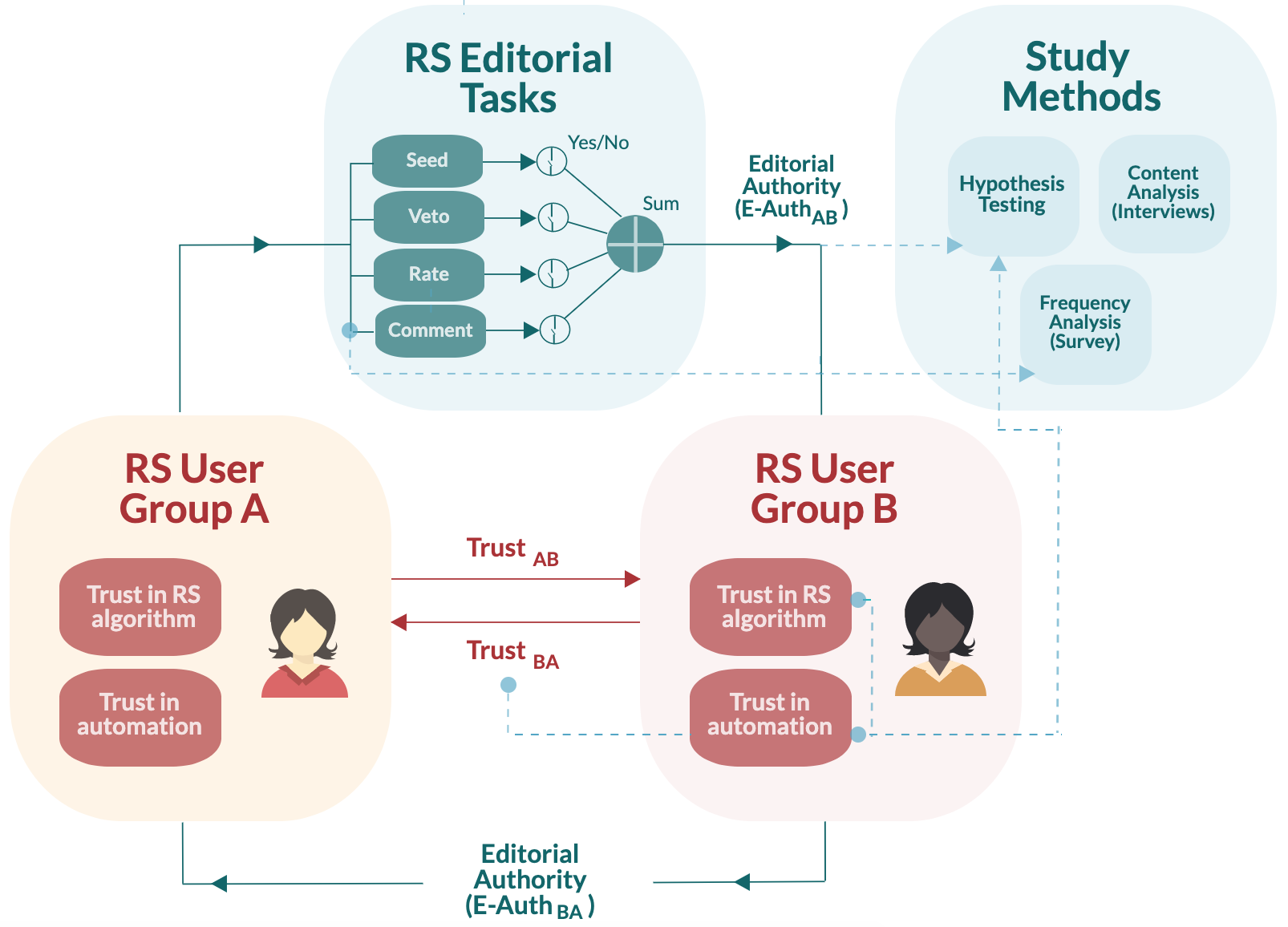}
  \caption{Study overview: recommender system (RS) user groups, trust relationships, attitudes about algorithms and automation, and mutual editorial role assignment. For our study, user groups include faculty, teaching assistants, and students.}
  \label{fig_sys_overview}
\end{figure*}

\section{Introduction}
Building user trust in recommendation systems (RS) is inextricably linked to concerns of algorithmic accountability and trust in automation. User experiments with RSs typically assess the \textit{experiential} notion of trust using the \textit{perceived} trustworthiness of the RS output. Algorithms for recommendation also frequently incorporate explicit and implicit signals of trust, for instance, the RS user's \textit{interactional} awareness of their local neighborhood, or some consensus of the preferences of their self-reported, trustworthy friends in a social network at large \cite{massa2007trust}\cite{hassan2019trust}\cite{hassan2019bias}. While algorithmic awareness of a user's neighborhood is important for producing accurate recommendations, real life recommendation tasks often involve user groups with differences in institutional, group, or task-based roles, powers and prerogatives. Building an educational RS, for example, needs us to model the interaction (or lack thereof) between teachers, teaching assistants and students in deriving value out of said RS. We argue that recognizing the editorial power relationships between stakeholders in the RS application domain is one way to begin to identify the broader context of trust in the recommendation algorithm, and to expand the interpretive power of the RS output. As a case study, we examine Virginia Tech faculty and students' preferences of editorial authority and trust in algorithms, for a hypothetical `Suggested Readings' recommender system aboard a learning management system (LMS) course site. Using a simple, first-principles metric of editorial authority (\textbf{E-Auth}), we hypothesis-test the relationship between RS editorial task distribution and stakeholder trust in algorithmic agency. We then describe the top three editorial roles (author, active viewer, viewer) allocated to students by course staff, and identify frequent themes, contexts and RS use cases using the trust relationships that define these editorial roles. Figure \ref{fig_sys_overview} provides an overview of our study methodology, and the group attitudes and trust relationships we investigate.

\begin{figure*}[t]
  \centering
  \includegraphics[width=1\linewidth]{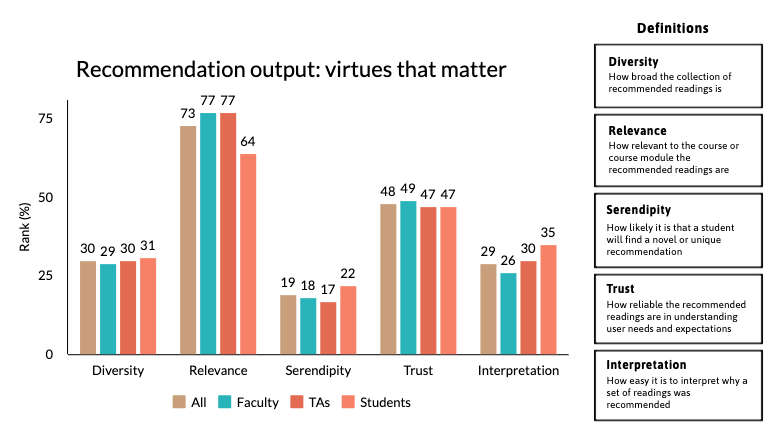}
  \caption{Virginia Tech faculty and students' perception of the relative importance of RS output virtues (with definitions). All three user roles (faculty, teaching assistants, students) rank relevance and trust consistently higher than interpretation, diversity and serendipity.}
  \label{fig_what_matters_more}
\end{figure*}

\section{Related Work}
There is ample prior work on both interpretive and algorithmic models of trust, in fields as diverse as recommender systems \cite{victor2011trust}, adaptive web systems \cite{artz2007survey}, user modeling \cite{guo2016social}\cite{pu2012evaluating}, computer networks \cite{kamvar2003eigentrust}, and game theory \cite{braynov2002contracting}. In a cross-domain review of trust frameworks, Artz and Gil \cite{artz2007survey} describe four broad paradigms of trust analyses (policy, reputation, information web, and general). The first three deal with algorithmic, heuristic, or policy-based views of trust inference for systems with human and software agents. For instance, Massa and Avesani \cite{massa2004trust}\cite{massa2007trust} develop an influential trust-aware recommendation framework which devises a \textit{local} trust propagation strategy: predicting the trustworthiness of the RS user's neighbors based on proximity to user-labeled trusted `peers'. A large body of work has been devoted to trust-aware collaborative filtering (CF), trust propagation, aggregation and user-similarity assessment using contextual, personal and network traits ever since \cite{shi2014collaborative}. Ma et al. \cite{ma2009learning}, for instance, propose an ensemble fusion of the ratings and web-of-trust similarity matrices. User experiments with recommender systems \cite{knijnenburg2015evaluating} often identify trust as one of the personal attributes of users or user groups. For instance, Knijnenburg et al. \cite{knijnenburg2011each} compare five interaction methods (topN, hybrid, explicit, implicit and hybrid) with a recommender system for energy saving measures. They discover that a user's propensity to trust is linked more to their choice satisfaction and trust in the overall recommendation system, rather than their preference for additional control or customization.

The final paradigm, in comparison, consists of sociological, psychological, and game-theoretic precursors and implications of trust. Mui et al. \cite{mui2002computational} propose a computational model using the notion of `reciprocity' (the bi-directional exchange of favors or revenge) to infer social reputation and trust. Braynov and Sandholm \cite{braynov2002contracting} discover that a misrepresentation of agents' trustworthiness or distrust can results in sub-optimal degrees of social welfare, profits and cooperation in a bilateral negotiation game. Note how aforementioned models often capture trust exchange between stakeholders, enhancing or undermining their \textit{reputation}, and by implication, trustworthiness of their social neighbors in the process. 

A parallel body of work on recommendation for groups \cite{gartrell2010enhancing}\cite{amer2009group} has grappled with the problem of evaluating group consensus when its various subgroups exhibit independent and overlapping preferences and roles. A subset of this research relies on social choice theory to find broad types of consensus (most popular items, least misery for group members, etc.) \cite{cantador2012group}. Seko et al. \cite{seko2011group} combine the notions of overlapping behavioral tendencies and power-balance in the group to assign consensus. Many open questions remain, including how to evaluate power asymmetry and assess the trustworthiness of algorithmic judgments of consensus. 

Our work seeks to advance a complementary and somewhat under-explored, actor network theory-inspired \cite{fenwick2010actor} approach to modeling the rich contexts in which educational RSs operate. We contend that editorial power assignment is a key early mechanism by which educational recommender systems are adopted by and appropriated for faculty, students and TAs to support teaching and research. Our approach assesses the distribution of trust preferences by RS editorial authority vested in these broad user roles. We then examine the distinctions between different stakeholders' power to source, veto, rate, and comment on recommended readings, and how their attitudes towards algorithmic agency and trust relationships with other stakeholders inform and affect these differences. We make the following contributions in this study.

\begin{enumerate}
    \item We propose a first-principles view of group trust in recommendation, based in editorial task assignment,
    
    \item We hypothesis-test the relationship between editorial authority (\textbf{E-Auth}) and trust in algorithmic agency,
    
    \item We infer key editorial roles allocated to students for educational recommendation, as well as assess staff needs and precedents that inform the corresponding trust relationships,
\end{enumerate}

\begin{table*}
  \caption{Survey questions: editorial authority and trust for a `Suggested Readings' recommender system}
  \label{tab:survey_questions}
  \begin{tabular}{lll}
    \toprule
    Question Category & Question Statement\\
    \midrule
    Editorial authority & Who can select seed articles for the RS? (Instructor, TA, student)\\
    & Who should have the `veto power'? (Instructor, TA, student)\\
     & Who should have the ability to rate the recommendations to influence their rank?\\
     & Who should be able to submit feedback on a recommendation?\\
    Relative trust in algorithms & In the short/long-term, who's responsible for trustworthy recommendations?\\
     & (Rank order: course staff, recommendation algorithm, students)\\
     Automation & In the short/long-term, who's responsible for trustworthy recommendations?\\
     & (Likert: all human - all algorithmic)\\
     Output attributes & What matters to you viz-a-viz RS output?\\
     & (Rank order: diversity, relevance, serendipity, trust, interpretation)\\
    \bottomrule
  \end{tabular}
\end{table*}

\begin{table*}[t]
  \caption{Key attributes of survey participants}
  \label{tab:keyattributes}
  \begin{tabular}{lcccc}
    \toprule
    Role & \# & \# Male/Female & \# Departments & \# STEM\\
    \midrule
    Faculty & 27 & 13/14 & 16 & 17\\
    Teaching assistant & 6 & 4/2 & 5 & 6\\
    Student & 9 & 7/2 & 4 & 7\\
    \bottomrule
  \end{tabular}
\end{table*}

\section{Editorial authority, trust and algorithmic agency} Recommender systems for education have fulfilled a variety of tasks for the individual learner, interpretation and intervention, in-class and online \cite{manouselis2011recommender}. However, there is a pressing need for the educational RS community to recognize platform-level changes in the domain. Recommender systems have to reckon with concerns of efficacy, trust and interpretation for \textit{multiple stakeholders}, and \textit{at scale} \cite{hassan2020depth}, especially as service-based learning management systems (LMS) become the primary infrastructure for productivity, communication and class management at institutions of higher learning \cite{mcgill2009task}. Figure \ref{fig_what_matters_more} describes five desirable aspects of RS output (relevance, trust, diversity, serendipity, and interpretation), and how faculty members and students who participated in our study felt about their relative importance. Relevance and trust were consistently ranked more important relative to the rest by all user roles. 

Our study investigates whether the differences in \textit{editorial authority} - exemplified in, say, an editor-consumer relationship between faculty and students - are related to the trust both assign to each other and the RS algorithm. It explores the most frequent RS editorial models preferred by the study participants, and their relationship with the trust exchanged between faculty, TAs, students and the RS algorithm.  For instance, faculty's willingness to incorporate student and TA feedback into the RS algorithm can point to a belief in editorial authority for multiple stakeholders, or regard for automation in the longer-term. Table \ref{tab:survey_questions} describes the tasks accounted for in editorial authority.

\subsection{Definitions}
We define the `Editorial Authority' (\textbf{E-Auth}) allocated to a study participant as a linear aggregate of all editorial powers (seed, veto, rate, comment) they identify for their own user group relative to all the other user groups. The individual editorial powers are represented by binary (`yes' or `no') votes. Consider equation 1 as follows.

\begin{equation}
\begin{aligned}
    \textbf{E-Auth (Faculty} \rightarrow \textbf{Students)} = \textbf{Seed Score} \: + \\ \textbf{Veto Score} + \textbf{Rate Score} + \textbf{Comment Score}
\end{aligned}
\end{equation}

For example, a faculty member's choice to let the TAs and students seed RS articles will result in a \textbf{seed score} of 2 (as in, the power to source recommended articles is shared with two user groups). If this faculty member favors exclusive veto power for faculty, but retains the rate/comment power for TAs and students, the \textbf{veto score}, \textbf{rate score} and \textbf{comment score} would be 0, 2 and 2 respectively, with a final \textbf{E-Auth} score of 6 (normalized to a percentage for our analyses). A small \textbf{E-Auth} score thus indicates that the course instructor might minimize student participation to reduce the cognitive effort of managing student feedback at scale, to prevent spam and inappropriate content, and to better align the course with learning outcomes in a given degree specialization. Section 5 uses frequent themes from semi-structured interviews to further discuss faculty rationales behind differing choices of editorial authority.

\subsection{Research questions and hypotheses}
In this section, we describe our study research questions which test the connection between editorial authority (\textbf{E-Auth}), trust in RS algorithm, and automation. 

\subsubsection{RQI: Trust in RS algorithm}
How does the relative trust in a `Suggested Readings' recommendation algorithm vary by job role (\textbf{RQ1a}), and how does it vary over time (\textbf{RQ1b})?

\subsubsection{RQII: Automation}
How much does the relative preference for algorithmic agency and control (over recommendation sourcing, updates and removal) vary by job role (\textbf{RQ2a}), and how does it vary over time (\textbf{RQ2b})?

\subsubsection{RQIII: Editorial authority, trust and automation}
What is the relationship between editorial authority assigned to students and the corresponding trust in RS algorithm (\textbf{RQ3a}) and preference for automation (\textbf{RQ3b})?

\section{Evaluation}
We conducted a small-scale, mixed-methods study to discover and interpret the Virginia Tech staff and students' perceived efficacy and trustworthiness of a hypothetical `Suggested Readings' RS. In this section, we describe the design, evaluation methods, key results, and limitations of the study. 

\begin{figure*}[t]
  \centering
  \includegraphics[width=0.99\linewidth]{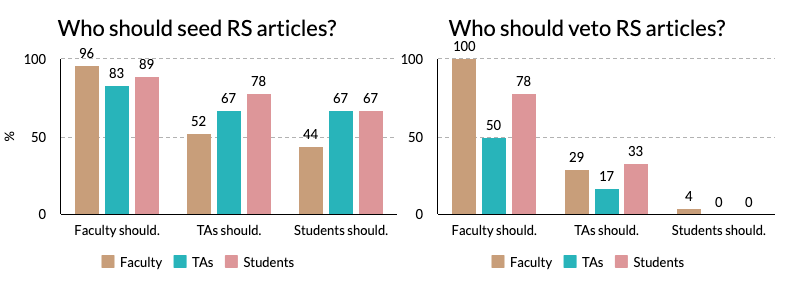}
  \caption{Editorial authority: Virginia Tech faculty, TA and students' opinions on who should be allowed to seed (left) and veto (right) articles for a `Suggested Readings' RS. 44\% of surveyed faculty allocated the article sourcing task to students, but only 4\% preferred students participate in article vetoing.}
  
  \label{fig:seed_and_veto}
\end{figure*}

\begin{figure*}[t]
  \centering
  \includegraphics[width=0.99\linewidth]{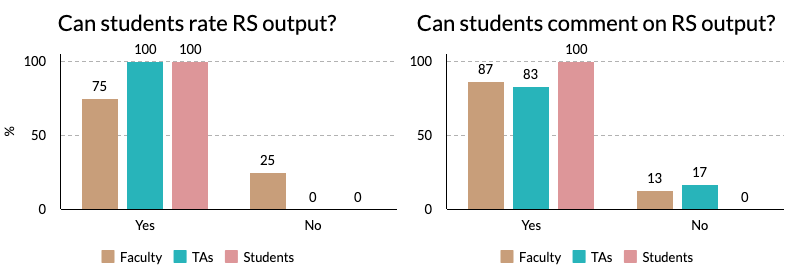}
  \caption{Editorial authority: Virginia Tech faculty, TA, and students' opinions on whether students should participate in rating and commenting on the output of a `Suggested Readings' RS. For instance, 25\% of surveyed faculty do not favor that students' ratings influence recommended articles, and 13\% do not favor student comments.}
  \label{fig:like_and_comment}
\end{figure*}

\subsection{Dataset and Methods}
Our dataset consists of survey responses from 42 participants (27 faculty, 6 teaching assistants, 9 students), and transcripts and notes from follow-up semi-structured interviews with 11 (6 faculty, 3 teaching assistants, 2 students) of the aforementioned 43, all affiliated with Virginia Tech. Table \ref{tab:keyattributes} details their key attributes. The faculty members in our survey represent 16 departments, while the TAs and students represent 5 and 4 departments respectively. 30 of the survey respondents (17 faculty members, 6 TAs, 7 students) represent STEM disciplines of study. We recruited participants on a rolling basis between August and October of 2020, using convenience sampling and voluntary response sampling on departmental mailing lists and Facebook groups. For our analyses, we use a mixed-methods approach. We perform hypothesis tests regarding the relationships between our study variables (relative trust in algorithms, automation, editorial authority) using one-way ANOVA (F-test). We also perform content analysis on survey responses and interview transcripts to identify frequent themes in the study participants' commentary regarding their preferred RS editorial tasks for each user role. 

\begin{table*}
  \caption{Key results from hypothesis tests about the relationship between \textbf{E-Auth} scores (RS editorial authority assigned by faculty to students) and staff/student attitudes. `Course staff' refers to faculty and teaching assistants considered together. Statistically significant effects ($F>F_{crit}, p<0.05$) are described in bold text in the rightmost column. For instance, per \textbf{H1}, teachers trust the RS algorithm less than teaching assistants, but according to  \textbf{H2}, course staff does not trust the RS algorithm any less than students do.}
  \label{tab:keyfacts}
  \begin{tabular}{p{1.5cm}p{1.2cm}lp{1.5cm}}
    \toprule
    Question Category & Research Question & Hypothesis & Effect sizes ($F,p$)\\
    \midrule
    Trust in RS & \textbf{RQ1a} & \textbf{H1}: Teachers trust the RS algorithm less than teaching assistants. & \textbf{6.49, 0.01}\\
    algorithm & \textbf{RQ1a} & \textbf{H2}: Course staff trusts the RS algorithm less than students. & 0.19, 0.66\\
     & \textbf{RQ1b} & \textbf{H3}: All participants trust the staff more short-term compared to long-term. & \textbf{10.8, <0.01}\\
     & \textbf{RQ1b} & \textbf{H4}: All participants trust the RS algorithm more long-term. & \textbf{6.3, 0.01}\\
    & & \\
    Automation & \textbf{RQ2a} & \textbf{H5}: Course staff trusts automation less in the short-term than students. & \textbf{9.7, <0.01}\\
     & \textbf{RQ2b} & \textbf{H6}: All respondents trust automation more long-term compared to short-term. & 3.9, 0.06\\
     & \textbf{RQ2b} & \textbf{H7}: Course staff trusts automation more long-term compared to short-term. & \textbf{7.2, <0.01}\\
    & & \\
    Editorial & \textbf{RQ3a} & \textbf{H8}: Higher trust in students long-term is linked to higher student \textbf{E-Auth}. & \textbf{3.7, 0.03}\\
    authority & \textbf{RQ3a} & \textbf{H9}: Lower average trust in RS algorithm is linked to higher student \textbf{E-Auth}. & \textbf{6.5, 0.01}\\
     & \textbf{RQ3b} & \textbf{H10}: Favoring high RS automation is linked to higher student \textbf{E-Auth}. & 0.52, 0.67\\
    \bottomrule
  \end{tabular}
\end{table*}

\subsection{Results}
Figure \ref{fig:seed_and_veto} illustrates how faculty, TAs and students prefer the distribution of RS tasks by job role. Nearly all faculty members allocate sourcing and vetoing tasks to themselves, but fewer (52\%) think TAs should seed RS articles and even fewer (44\%) allocate this task to students. The divide is quite apparent for vetoing, as all but one (4\%) faculty member suggest that students should not be able to instantly remove recommended readings. As per figure \ref{fig:like_and_comment}, about 75\% of faculty (as opposed to all students) favor the idea of soft power for students: the ability to rate individual recommendations and let the algorithm update the relative location of a recommended reading using group consensus strategies. 13\% of faculty members and 17\% of teaching assistants suggest that they do not favor students' ability to submit feedback about a recommended reading.

Table \ref{tab:keyfacts} describes the results of hypothesis tests on key claims about trust in RS algorithm, preference for automation, and aggregate editorial authority (see research questions \textbf{RQ1-3} in section 3.2). We discover that overall, course staff (faculty and teaching assistants) tends to be more risk-averse in its trust relationship with RS algorithm and automation than students, favoring substantial human intervention in the sourcing, updating and removal of recommended readings in the short-term. Here is a detailed breakdown of these results.

\subsubsection{Trust in RS algorithm (\textbf{RQ1}, \textbf{H1}-\textbf{H4})}
We discover that teachers tend to trust the RS algorithm less than teaching assistants ($F=6.49, p=0.01$). All user groups favor the role of course staff in ensuring trustworthy recommendations short-term more than long-term ($F=10.8, p=1e^{-4}$). Similarly, all user groups trust the RS algorithm more long-term compared to short-term ($F=6.3, p=0.01$). 

\subsubsection{Automation (\textbf{RQ2}, \textbf{H5}-\textbf{H7})} Course staff tends to favor automation in recommendations less relative to students, especially in the short-term ($F=9.7,p=3e^{-3}$). Similarly, course staff favors automation in the long-term more than in the short-term ($F=7.2,p=9e^{-3}$).

\subsubsection{Editorial authority (\textbf{RQ3}, \textbf{H8}-\textbf{H10})}
Higher trust in students is linked to preference for higher student \textbf{E-Auth} scores in the long term ($F=3.7, p=0.03$), hinting at a connection between perceived trust and editorial role assignment. Similarly, lower average trust in RS algorithm is linked to preference for higher student \textbf{E-Auth} scores ($F=6.5, p=0.01$). Preference for higher RS automation overall is however, not linked to editorial authority ($F=0.52, p=0.67$).

\section{Discussion and Lessons Learned}
In this section, we describe the top three frequently recommended editorial roles (author, active viewer, viewer) for students, as allocated by our study respondents (table \ref{tab:editorialpowerallocation}). Allocation refers to the fraction (\%) of participants who chose to allocate a given editorial role (and corresponding powers and tasks) to students. For each editorial role, we discuss the frequent themes (in bold text) of the mediating trust relationships we investigate in \textbf{RQs I} through \textbf{III}. These emerge from our content analysis, and they include, and are not limited to, differences in perceived expertise, supporting learner engagement, and scaling the learning environment. These editorial roles highlight distinct - if overlapping - user preferences that can suggest RS use-cases for supporting both teaching and research.

\begin{table*}[t]
  \caption{RS editorial roles allocated to students. Editorial tasks essential to each role are in bold text. Allocation refers to the number and fraction (\%) of participants who chose to allocate a given editorial role to students. Average \textbf{E-Auth} refers to the amount of editorial authority assigned to students in each role. For instance, 39\% of course staff favored students provide input as active viewers (AV), assigning them 54\% of the available editorial authority.}
  \label{tab:editorialpowerallocation}
  \begin{tabular}{lllll}
    \toprule
    Editorial Role & Editorial Powers/Tasks & Average \textbf{E-Auth} & Overall Allocation & Staff Allocation \\
    \midrule
    Editor (\textbf{E}) & \textbf{Seed}, \textbf{veto}, rate, comment  & 80\% & 1 (2\%) & 1 (3\%)\\
    Author (\textbf{A}) & \textbf{Seed}, rate, comment  & 77\% & 17 (40\%) & 12 (36\%)\\
    Active Viewer (\textbf{AV}) & \textbf{Rate}, comment  & 54\% & 17 (40\%) & 13 (39\%)\\
    Viewer (\textbf{V}) & Comment & 14\% & 7 (16\%) & 7 (21\%)\\
    \bottomrule
  \end{tabular}
\end{table*}

\subsection{"The author" (\textbf{A}): Everything except veto}
According to table \ref{tab:editorialpowerallocation},  40\% of participants overall, and 36\% of course staff favored the `author' (\textbf{A}) role for students. \textbf{A}-faculty members prefer sharing article sourcing, rating and feedback responsibilities with students, but retain the veto power (i.e. the ability to instantly remove a recommended reading for all RS users) for the primary course instructor, as well as for the teaching assistants in a subset of cases. Table \ref{tab:frequentthemes} lists the frequent themes and course contexts for all RS editorial roles. In addition to citing their prerogative being in charge of facilitating student learning, \textbf{A}-faculty members frequently cited the \textbf{need to moderate content} and remove readings deemed irrelevant or malicious. According to one instructor: "\textit{I worry about abusive behavior, unless it was clearly tracked}" and this would "\textit{ensure the integrity of the course and the suggested readings}". They simultaneously acknowledged the potential utility of \textbf{student engagement} in these editorial tasks, and inquired about the ability of the RS algorithm to comprehend student needs early on (\textbf{H1}, \textbf{H4}), especially in graduate, research-focused courses. One faculty member said that based on her teaching experience, she perceived MS-level students to be resourceful and another commented: 

\begin{quote}"\textit{.. if objectionable material was posted, I'd rather we use it as a discussion topic rather than outright reject it.}"\end{quote}

Soft power, as in the ability to rate (or `like' and `dislike') content as a signal to RS algorithm, was frequently favored for students as a means of further engaging them with the course content. An instructor said of soft power, \textit{"I like the way the soft power idea will engage their minds and (aid their) learning"}. This is consistent with hypothesis \textbf{H8} and \textbf{H9} in table \ref{tab:keyfacts}. Teaching assistants often brought up the case of large course sections with multiple instructors, and how TAs and students having article sourcing powers could potentially help scale the teaching resources faster, and make them accessible to a larger subset of the student population. According to one teaching assistant,

\begin{quote}"\textit{.. course instructors may sometimes not know when to remove content unless they are intimately familiar with it and know it to not be useful.}"\end{quote}

Student participants in our study were, by-and-large, in agreement with course staff about exclusive veto power for faculty and teaching assistants, especially because they felt that the ability to seed, rate and comment on articles gave them ample opportunity to engage with the course materials and RS authorship policies. As per one student: \textit{"I feel anything that is against the teaching of the class should be allowed and a respectful discussion should occur; preferably in a forum setting."} 

\begin{table*}[t]
  \caption{Frequent course contexts, themes, and use-cases for RS editorial roles allocated to students. For instance, study participants favored cooperative RS editorial models (\textbf{E}/\textbf{A}) for small class sizes (typically graduate courses). Courses with a preference for lesser or no student input (\textbf{AV}/\textbf{V}) have large class sizes (often undergraduate) and time constraints defined by course content (STEM, professional diplomas).}
  \label{tab:frequentthemes}
  \begin{tabular}{clll}
    \toprule
    Editorial Role & Frequent Course Context & Frequent Themes & RS Use Cases\\
    \midrule
    
    \textbf{E/A} & Graduate, Non-STEM & student engagement, content & teaching\\
    & & moderation &\\
    \textbf{AV} & Undergraduate, STEM & student disengagement, scale, & teaching, professional\\
    & & outcome bias, content moderation & development\\
    \textbf{V} & Undergraduate, STEM & student disengagement, scale, & research, teaching\\
    & & content moderation & \\
    
    \bottomrule
  \end{tabular}
\end{table*}

\subsection{"The active viewer" (\textbf{AV}): Rate and comment}
Table \ref{tab:editorialpowerallocation} suggests that 40\% of study participants overall, and 39\% of course staff favor an `active viewer' role for students. This allocates the RS rating and commenting tasks to students while reserving the sourcing and vetoing of suggested readings for course staff. Members of this group frequently talk about the challenges of managing and responding to feedback at \textbf{scale}, especially in undergraduate courses with multiple sections and hundreds of students. One faculty member remarked about students not being able to seed or veto suggested readings:

\begin{quote}"\textit{.. (this is) just to be able to manage with a course that has 14-15 sections of 65 students each.}"\end{quote}

This comment is not a lone anomaly. Over the last two years, the average undergraduate course at Virginia Tech has 3.7X the class size of the average graduate course (66 and 17 students on average, respectively), with many first-year, major-unrestricted courses enrolling several hundred students in any given term. It is worthwhile noting that \textbf{AV}-faculty members do not seek a fully cooperative model of student feedback to solve this challenge in the manner of \textbf{A}-faculty. Frequent reasons posited by faculty include perceived historical patterns of \textbf{student disengagement} and exclusive attention by students to course outcomes, or the \textbf{outcome bias} \cite{hassan2019bias}. This bias is also known to affect student perceptions of instructors on academic social forums like \textit{Koofers} \cite{koofershomepage} and \textit{RateMyProfessor} \cite{rmphomepage}. As one faculty member commented, 

\begin{quote}"\textit{My experience has been that most students do not do more than what's required of them, unfortunately. They just want to know what they can do to earn an A in class.}"\end{quote} 

Another faculty member talked about the requirements of professional diplomas - dictated by the rapidly evolving demands of the job market - as one key driver of student disengagement. He noted that the "\textit{fast pace and skill-focused curriculum}" of professional diplomas (as opposed to research-based degrees) made it difficult for students to spend time on optional course content. In his experience, this had led to a substantial decline in student interaction with the course LMS site, to the point where cooperative editing of recommended readings seemed ineffective. In comparison, teaching assistants and students who favored the \textbf{AV} role largely cited course staff's \textbf{need for moderating} the RS articles for malicious behavior and honor code violations. About the article rating task, one student commented that it will "\textit{allow the instructor and TAs to see what's most widely accepted}". Consistent with hypothesis \textbf{H7} (table \ref{tab:keyfacts}), all participants expected the role of automation to be significantly higher long-term in consolidating student updates to the recommendations' rank order. 

\subsection{"The viewer" (\textbf{V}): Comment only}
Study participants in this group favored the least egalitarian editorial model for recommending readings to students. 16\% of participants overall, and 21\% of faculty and teaching assistants prefer no direct input from students in deciding the source and rank order of recommended readings. Same as \textbf{AV}, this is frequently observed for faculty members with undergraduate teaching responsibilities. This model considers student participation in article sourcing, updates and removal to be unsustainable if not counter-productive. Faculty survey respondents and interviewees frequently cited challenges of \textbf{content moderation} at \textbf{scale}. One instructor of an undergraduate Computer Science class described the problem of cooperative RS editing as analogous to the challenge of regulating discussion forums posts for the 200+ enrolled students in her section. Drawing on instances of age-sensitive commentary by students about an automated grading software, this instructor suggested that comments on recommended readings be invisible to fellow students by default. This would allow course staff to remove malicious content and notify the students. Another course instructor in this group complained "\textit{.. unfortunately, students are often testing the limits of the honor code.}".

An Engineering instructor cited his time in the industry as having informed his singular emphasis on problem-solving in teaching graduate courses. He mentioned he favored creating and updating course assignments, projects, and exams without serious consultation from a primary textbook. This, according to him, discouraged students' use of online solution manuals, better assessed their progress with course milestones, and ensured they learned a precise set of Engineering skills without \textbf{undue cognitive burden}. Note the parallel with \textbf{AV}-faculty's rationale for the limited role of a `Suggested Readings' RS in teaching courses for professional diplomas. About restricting students to a comment-only RS editorial role, he commented: 

\begin{quote}"\textit{It's not so much that I want the control. It's more that I don't think that should be there focus from an educational standpoint. I'm trying to get them to comprehend some pretty intense Engineering and design concepts. I consider it a burden for them to go out and find other resources. I really have to get them to practice a lot of things over and over.}"\end{quote}

As per table \ref{tab:frequentthemes}, several \textbf{V}-faculty suggested they might use the recommender system in \textbf{supporting research}, as opposed to assorting readings for teaching purposes. The frequently cited use-cases for such a recommender system were discovering research articles beyond Virginia Tech library-indexed databases, and discovering topic overlap with other research fields to inform literature reviews. It is worth noting that no teaching assistants or students favored a \textbf{V}-role for students, hinting at a strongly asymmetric editorial relationship between faculty and students. A more rigorous evaluation of this power asymmetry is left for future work.

\subsection{Editor-consumer relationships}
We observe frequent power asymmetry when it comes to editorial preferences for recommendation in the education domain. 25\% and 13\% of faculty members surveyed did not favor students' ability to rate and comment on the recommended readings, respectively (figure \ref{fig:like_and_comment}). As we observe earlier in this section, this frequently coincides with complaints of inevitable learner disengagement with course materials, but the course staff's comfort with automating and delegating editorial tasks increases with time. Note the connection with passive use of recommender systems observed, for instance, in music and video recommenders where the recommendation algorithm has disproportionate editorial power (both explicit and implicit) as a function of time relative to passive consumers. While outside the scope of this study, it is important to investigate the UX implications of such a disparity using trust frameworks that take into account the capacity of the recommendation system to incorporate context \cite{hasan2017collaborative}, increase meaningful use through diverse use-cases anchored in trust, provide a diverse set of interaction techniques \cite{knijnenburg2015evaluating}, and disrupt filter bubbles \cite{nagulendra2014understanding}.

\section{Conclusions and Future Work}
We examine the distribution of editorial preferences and its broader roots in notions of trust, expertise and institutional and domain-specific prerogatives. We discover that power asymmetry is an important component of the user experience of educational recommender systems, and should be leveraged in algorithmic and interactional affordances for trust-aware recommender systems at scale.

The scope of our pilot study can be extended in several important ways. We review these as directions for future work as follows. The sparsity of interview evidence at this stage of the project makes it difficult to generalize our conclusions across the Virginia Tech faculty and student population at large. We reached out to course staff with a history of collaboration with departmental initiatives towards instructional design. For our future work, we would like to collect a comprehensive sample of faculty members, especially the ones who need to prioritize research over teaching, favor legacy tools, or experience a high cognitive burden-of-discovery \cite{west2007understanding} when adopting new digital tools. Given the variability of trust preferences is itself different for faculty, students and teaching assistants, we also plan to incorporate more robust multi-item trust and automation scales. We also intend to pursue a concrete definition of present bias in RS users, in order to thoroughly interpret the time variance of editorial preferences we observed in our study. Concrete experimental conditions with RS algorithm strata (learning strategies, interaction methods) and domain-specific timescales (semester, course project timeline, course add/drop/withdraw periods) would further improve the consistency and actionability of our conclusions. Finally, we plan to measure the actual use patterns of our personalized `Suggested Readings' recommendation system as a function of our inferred editorial roles. It is part of a larger effort by our team to understand the adoption of an institution-wide learning management system (LMS) and its managed apps. LMS adoption is a complex function of department-level precedents, faculty's cognitive burden-of-discovery, task-technology fit, as well as the needs for scale and ubiquitous access \cite{coates2005critical}. In our future work, we hope to expand the interpretive power of LMS platform analytics using our multistakeholder trust models.



\bibliographystyle{ACM-Reference-Format}
\bibliography{sample-base}


\begin{thebibliography}{29}


\ifx \showCODEN    \undefined \def \showCODEN     #1{\unskip}     \fi
\ifx \showDOI      \undefined \def \showDOI       #1{#1}\fi
\ifx \showISBNx    \undefined \def \showISBNx     #1{\unskip}     \fi
\ifx \showISBNxiii \undefined \def \showISBNxiii  #1{\unskip}     \fi
\ifx \showISSN     \undefined \def \showISSN      #1{\unskip}     \fi
\ifx \showLCCN     \undefined \def \showLCCN      #1{\unskip}     \fi
\ifx \shownote     \undefined \def \shownote      #1{#1}          \fi
\ifx \showarticletitle \undefined \def \showarticletitle #1{#1}   \fi
\ifx \showURL      \undefined \def \showURL       {\relax}        \fi
\providecommand\bibfield[2]{#2}
\providecommand\bibinfo[2]{#2}
\providecommand\natexlab[1]{#1}
\providecommand\showeprint[2][]{arXiv:#2}

\bibitem[\protect\citeauthoryear{??}{rmp}{2019}]%
        {rmphomepage}
 \bibinfo{year}{2019}\natexlab{}.
\newblock \bibinfo{booktitle}{\emph{Rate My Professors - review teachers and
  professors, school reviews, college campus ratings}}.
\newblock
\urldef\tempurl%
\url{http://ratemyprofessors.com/}
\showURL{%
Retrieved 2019 from \tempurl}


\bibitem[\protect\citeauthoryear{Amer-Yahia, Roy, Chawlat, Das, and
  Yu}{Amer-Yahia et~al\mbox{.}}{2009}]%
        {amer2009group}
\bibfield{author}{\bibinfo{person}{Sihem Amer-Yahia},
  \bibinfo{person}{Senjuti~Basu Roy}, \bibinfo{person}{Ashish Chawlat},
  \bibinfo{person}{Gautam Das}, {and} \bibinfo{person}{Cong Yu}.}
  \bibinfo{year}{2009}\natexlab{}.
\newblock \showarticletitle{Group recommendation: Semantics and efficiency}.
\newblock \bibinfo{journal}{\emph{Proceedings of the VLDB Endowment}}
  \bibinfo{volume}{2}, \bibinfo{number}{1} (\bibinfo{year}{2009}),
  \bibinfo{pages}{754--765}.
\newblock


\bibitem[\protect\citeauthoryear{Artz and Gil}{Artz and Gil}{2007}]%
        {artz2007survey}
\bibfield{author}{\bibinfo{person}{Donovan Artz} {and} \bibinfo{person}{Yolanda
  Gil}.} \bibinfo{year}{2007}\natexlab{}.
\newblock \showarticletitle{A survey of trust in computer science and the
  semantic web}.
\newblock \bibinfo{journal}{\emph{Journal of Web Semantics}}
  \bibinfo{volume}{5}, \bibinfo{number}{2} (\bibinfo{year}{2007}),
  \bibinfo{pages}{58--71}.
\newblock


\bibitem[\protect\citeauthoryear{Braynov and Sandholm}{Braynov and
  Sandholm}{2002}]%
        {braynov2002contracting}
\bibfield{author}{\bibinfo{person}{Sviatoslav Braynov} {and}
  \bibinfo{person}{Tuomas Sandholm}.} \bibinfo{year}{2002}\natexlab{}.
\newblock \showarticletitle{Contracting with uncertain level of trust}.
\newblock \bibinfo{journal}{\emph{Computational Intelligence}}
  \bibinfo{volume}{18}, \bibinfo{number}{4} (\bibinfo{year}{2002}),
  \bibinfo{pages}{501--514}.
\newblock


\bibitem[\protect\citeauthoryear{Cantador and Castells}{Cantador and
  Castells}{2012}]%
        {cantador2012group}
\bibfield{author}{\bibinfo{person}{Iv{\'a}n Cantador} {and}
  \bibinfo{person}{Pablo Castells}.} \bibinfo{year}{2012}\natexlab{}.
\newblock \showarticletitle{Group recommender systems: new perspectives in the
  social web}.
\newblock In \bibinfo{booktitle}{\emph{Recommender systems for the social
  web}}. \bibinfo{publisher}{Springer}, \bibinfo{pages}{139--157}.
\newblock


\bibitem[\protect\citeauthoryear{Coates, James, and Baldwin}{Coates
  et~al\mbox{.}}{2005}]%
        {coates2005critical}
\bibfield{author}{\bibinfo{person}{Hamish Coates}, \bibinfo{person}{Richard
  James}, {and} \bibinfo{person}{Gabrielle Baldwin}.}
  \bibinfo{year}{2005}\natexlab{}.
\newblock \showarticletitle{A critical examination of the effects of learning
  management systems on university teaching and learning}.
\newblock \bibinfo{journal}{\emph{Tertiary education and management}}
  \bibinfo{volume}{11} (\bibinfo{year}{2005}), \bibinfo{pages}{19--36}.
\newblock


\bibitem[\protect\citeauthoryear{Fenwick and Edwards}{Fenwick and
  Edwards}{2010}]%
        {fenwick2010actor}
\bibfield{author}{\bibinfo{person}{Tara Fenwick} {and} \bibinfo{person}{Richard
  Edwards}.} \bibinfo{year}{2010}\natexlab{}.
\newblock \bibinfo{booktitle}{\emph{Actor-network theory in education}}.
\newblock \bibinfo{publisher}{Routledge}.
\newblock


\bibitem[\protect\citeauthoryear{Gartrell, Xing, Lv, Beach, Han, Mishra, and
  Seada}{Gartrell et~al\mbox{.}}{2010}]%
        {gartrell2010enhancing}
\bibfield{author}{\bibinfo{person}{Mike Gartrell}, \bibinfo{person}{Xinyu
  Xing}, \bibinfo{person}{Qin Lv}, \bibinfo{person}{Aaron Beach},
  \bibinfo{person}{Richard Han}, \bibinfo{person}{Shivakant Mishra}, {and}
  \bibinfo{person}{Karim Seada}.} \bibinfo{year}{2010}\natexlab{}.
\newblock \showarticletitle{Enhancing group recommendation by incorporating
  social relationship interactions}. In \bibinfo{booktitle}{\emph{Proceedings
  of the 16th ACM international conference on Supporting group work}}.
  \bibinfo{pages}{97--106}.
\newblock


\bibitem[\protect\citeauthoryear{Glynn~LoPresti and Le}{Glynn~LoPresti and
  Le}{2019}]%
        {koofershomepage}
\bibfield{author}{\bibinfo{person}{Dan~Donahoe Glynn~LoPresti, Patrick~Gartlan}
  {and} \bibinfo{person}{Minhe Le}.} \bibinfo{year}{2019}\natexlab{}.
\newblock \bibinfo{booktitle}{\emph{Koofers - professor ratings, practice exams
  and flash cards}}.
\newblock
\urldef\tempurl%
\url{http://koofers.com/}
\showURL{%
Retrieved 2019 from \tempurl}


\bibitem[\protect\citeauthoryear{Guo, Zhu, Li, Wang, and Han}{Guo
  et~al\mbox{.}}{2016}]%
        {guo2016social}
\bibfield{author}{\bibinfo{person}{Junpeng Guo}, \bibinfo{person}{Yanlin Zhu},
  \bibinfo{person}{Aiai Li}, \bibinfo{person}{Qipeng Wang}, {and}
  \bibinfo{person}{Weiguo Han}.} \bibinfo{year}{2016}\natexlab{}.
\newblock \showarticletitle{A social influence approach for group user modeling
  in group recommendation systems}.
\newblock \bibinfo{journal}{\emph{IEEE Intelligent systems}}
  \bibinfo{volume}{31}, \bibinfo{number}{5} (\bibinfo{year}{2016}),
  \bibinfo{pages}{40--48}.
\newblock


\bibitem[\protect\citeauthoryear{Hasan, Arshad, Dahlquist, and
  McCrickard}{Hasan et~al\mbox{.}}{2017}]%
        {hasan2017collaborative}
\bibfield{author}{\bibinfo{person}{Taha Hasan}, \bibinfo{person}{Naveed
  Arshad}, \bibinfo{person}{Erik Dahlquist}, {and} \bibinfo{person}{Scott
  McCrickard}.} \bibinfo{year}{2017}\natexlab{}.
\newblock \showarticletitle{Collaborative filtering for household load
  prediction given contextual information}. In
  \bibinfo{booktitle}{\emph{Proceedings of the 2017 SIAM Workshop on Machine
  Learning for Recommender Systems (MLRec)}}.
\newblock


\bibitem[\protect\citeauthoryear{Hassan}{Hassan}{2019a}]%
        {hassan2019bias}
\bibfield{author}{\bibinfo{person}{Taha Hassan}.}
  \bibinfo{year}{2019}\natexlab{a}.
\newblock \showarticletitle{On bias in social reviews of university courses}.
  In \bibinfo{booktitle}{\emph{Companion Publication of the 10th ACM Conference
  on Web Science}}. \bibinfo{pages}{11--14}.
\newblock


\bibitem[\protect\citeauthoryear{Hassan}{Hassan}{2019b}]%
        {hassan2019trust}
\bibfield{author}{\bibinfo{person}{Taha Hassan}.}
  \bibinfo{year}{2019}\natexlab{b}.
\newblock \showarticletitle{Trust and trustworthiness in social recommender
  systems}. In \bibinfo{booktitle}{\emph{Companion Proceedings of The 2019
  World Wide Web Conference}}. \bibinfo{pages}{529--532}.
\newblock


\bibitem[\protect\citeauthoryear{Hassan, Edmison, Cox, Louvet, Williams, and
  McCrickard}{Hassan et~al\mbox{.}}{2020}]%
        {hassan2020depth}
\bibfield{author}{\bibinfo{person}{Taha Hassan}, \bibinfo{person}{Bob Edmison},
  \bibinfo{person}{Larry Cox}, \bibinfo{person}{Matt Louvet},
  \bibinfo{person}{Daron Williams}, {and} \bibinfo{person}{D~Scott
  McCrickard}.} \bibinfo{year}{2020}\natexlab{}.
\newblock \showarticletitle{Depth of Use: An Empirical Framework to Help
  Faculty Gauge the Relative Impact of Learning Management System Tools}. In
  \bibinfo{booktitle}{\emph{Proceedings of the 2020 ACM Conference on
  Innovation and Technology in Computer Science Education}}.
  \bibinfo{pages}{47--53}.
\newblock


\bibitem[\protect\citeauthoryear{Kamvar, Schlosser, and Garcia-Molina}{Kamvar
  et~al\mbox{.}}{2003}]%
        {kamvar2003eigentrust}
\bibfield{author}{\bibinfo{person}{Sepandar~D Kamvar}, \bibinfo{person}{Mario~T
  Schlosser}, {and} \bibinfo{person}{Hector Garcia-Molina}.}
  \bibinfo{year}{2003}\natexlab{}.
\newblock \showarticletitle{The eigentrust algorithm for reputation management
  in p2p networks}. In \bibinfo{booktitle}{\emph{Proceedings of the 12th
  international conference on World Wide Web}}. \bibinfo{pages}{640--651}.
\newblock


\bibitem[\protect\citeauthoryear{Knijnenburg, Reijmer, and
  Willemsen}{Knijnenburg et~al\mbox{.}}{2011}]%
        {knijnenburg2011each}
\bibfield{author}{\bibinfo{person}{Bart~P Knijnenburg},
  \bibinfo{person}{Niels~JM Reijmer}, {and} \bibinfo{person}{Martijn~C
  Willemsen}.} \bibinfo{year}{2011}\natexlab{}.
\newblock \showarticletitle{Each to his own: how different users call for
  different interaction methods in recommender systems}. In
  \bibinfo{booktitle}{\emph{Proceedings of the fifth ACM conference on
  Recommender systems}}. \bibinfo{pages}{141--148}.
\newblock


\bibitem[\protect\citeauthoryear{Knijnenburg and Willemsen}{Knijnenburg and
  Willemsen}{2015}]%
        {knijnenburg2015evaluating}
\bibfield{author}{\bibinfo{person}{Bart~P Knijnenburg} {and}
  \bibinfo{person}{Martijn~C Willemsen}.} \bibinfo{year}{2015}\natexlab{}.
\newblock \showarticletitle{Evaluating recommender systems with user
  experiments}.
\newblock In \bibinfo{booktitle}{\emph{Recommender Systems Handbook}}.
  \bibinfo{publisher}{Springer}, \bibinfo{pages}{309--352}.
\newblock


\bibitem[\protect\citeauthoryear{Ma, King, and Lyu}{Ma et~al\mbox{.}}{2009}]%
        {ma2009learning}
\bibfield{author}{\bibinfo{person}{Hao Ma}, \bibinfo{person}{Irwin King}, {and}
  \bibinfo{person}{Michael~R Lyu}.} \bibinfo{year}{2009}\natexlab{}.
\newblock \showarticletitle{Learning to recommend with social trust ensemble}.
  In \bibinfo{booktitle}{\emph{Proceedings of the 32nd international ACM SIGIR
  conference on Research and development in information retrieval}}.
  \bibinfo{pages}{203--210}.
\newblock


\bibitem[\protect\citeauthoryear{Manouselis, Drachsler, Vuorikari, Hummel, and
  Koper}{Manouselis et~al\mbox{.}}{2011}]%
        {manouselis2011recommender}
\bibfield{author}{\bibinfo{person}{Nikos Manouselis}, \bibinfo{person}{Hendrik
  Drachsler}, \bibinfo{person}{Riina Vuorikari}, \bibinfo{person}{Hans Hummel},
  {and} \bibinfo{person}{Rob Koper}.} \bibinfo{year}{2011}\natexlab{}.
\newblock \showarticletitle{Recommender systems in technology enhanced
  learning}.
\newblock In \bibinfo{booktitle}{\emph{Recommender systems handbook}}.
  \bibinfo{publisher}{Springer}, \bibinfo{pages}{387--415}.
\newblock


\bibitem[\protect\citeauthoryear{Massa and Avesani}{Massa and Avesani}{2004}]%
        {massa2004trust}
\bibfield{author}{\bibinfo{person}{Paolo Massa} {and} \bibinfo{person}{Paolo
  Avesani}.} \bibinfo{year}{2004}\natexlab{}.
\newblock \showarticletitle{Trust-aware collaborative filtering for recommender
  systems}. In \bibinfo{booktitle}{\emph{OTM Confederated International
  Conferences" On the Move to Meaningful Internet Systems"}}. Springer,
  \bibinfo{pages}{492--508}.
\newblock


\bibitem[\protect\citeauthoryear{Massa and Avesani}{Massa and Avesani}{2007}]%
        {massa2007trust}
\bibfield{author}{\bibinfo{person}{Paolo Massa} {and} \bibinfo{person}{Paolo
  Avesani}.} \bibinfo{year}{2007}\natexlab{}.
\newblock \showarticletitle{Trust-aware recommender systems}. In
  \bibinfo{booktitle}{\emph{Proceedings of the 2007 ACM conference on
  Recommender systems}}. \bibinfo{pages}{17--24}.
\newblock


\bibitem[\protect\citeauthoryear{McGill and Klobas}{McGill and Klobas}{2009}]%
        {mcgill2009task}
\bibfield{author}{\bibinfo{person}{Tanya~J McGill} {and}
  \bibinfo{person}{Jane~E Klobas}.} \bibinfo{year}{2009}\natexlab{}.
\newblock \showarticletitle{A task--technology fit view of learning management
  system impact}.
\newblock \bibinfo{journal}{\emph{Computers \& Education}}
  \bibinfo{volume}{52}, \bibinfo{number}{2} (\bibinfo{year}{2009}),
  \bibinfo{pages}{496--508}.
\newblock


\bibitem[\protect\citeauthoryear{Mui, Mohtashemi, and Halberstadt}{Mui
  et~al\mbox{.}}{2002}]%
        {mui2002computational}
\bibfield{author}{\bibinfo{person}{Lik Mui}, \bibinfo{person}{Mojdeh
  Mohtashemi}, {and} \bibinfo{person}{Ari Halberstadt}.}
  \bibinfo{year}{2002}\natexlab{}.
\newblock \showarticletitle{A computational model of trust and reputation}. In
  \bibinfo{booktitle}{\emph{Proceedings of the 35th Annual Hawaii International
  Conference on System Sciences}}. IEEE, \bibinfo{pages}{2431--2439}.
\newblock


\bibitem[\protect\citeauthoryear{Nagulendra and Vassileva}{Nagulendra and
  Vassileva}{2014}]%
        {nagulendra2014understanding}
\bibfield{author}{\bibinfo{person}{Sayooran Nagulendra} {and}
  \bibinfo{person}{Julita Vassileva}.} \bibinfo{year}{2014}\natexlab{}.
\newblock \showarticletitle{Understanding and controlling the filter bubble
  through interactive visualization: a user study}. In
  \bibinfo{booktitle}{\emph{Proceedings of the 25th ACM conference on Hypertext
  and social media}}. \bibinfo{pages}{107--115}.
\newblock


\bibitem[\protect\citeauthoryear{Pu, Chen, and Hu}{Pu et~al\mbox{.}}{2012}]%
        {pu2012evaluating}
\bibfield{author}{\bibinfo{person}{Pearl Pu}, \bibinfo{person}{Li Chen}, {and}
  \bibinfo{person}{Rong Hu}.} \bibinfo{year}{2012}\natexlab{}.
\newblock \showarticletitle{Evaluating recommender systems from the user’s
  perspective: survey of the state of the art}.
\newblock \bibinfo{journal}{\emph{User Modeling and User-Adapted Interaction}}
  \bibinfo{volume}{22}, \bibinfo{number}{4-5} (\bibinfo{year}{2012}),
  \bibinfo{pages}{317--355}.
\newblock


\bibitem[\protect\citeauthoryear{Seko, Yagi, Motegi, and Muto}{Seko
  et~al\mbox{.}}{2011}]%
        {seko2011group}
\bibfield{author}{\bibinfo{person}{Shunichi Seko}, \bibinfo{person}{Takashi
  Yagi}, \bibinfo{person}{Manabu Motegi}, {and} \bibinfo{person}{Shinyo Muto}.}
  \bibinfo{year}{2011}\natexlab{}.
\newblock \showarticletitle{Group recommendation using feature space
  representing behavioral tendency and power balance among members}. In
  \bibinfo{booktitle}{\emph{Proceedings of the fifth ACM conference on
  Recommender systems}}. \bibinfo{pages}{101--108}.
\newblock


\bibitem[\protect\citeauthoryear{Shi, Larson, and Hanjalic}{Shi
  et~al\mbox{.}}{2014}]%
        {shi2014collaborative}
\bibfield{author}{\bibinfo{person}{Yue Shi}, \bibinfo{person}{Martha Larson},
  {and} \bibinfo{person}{Alan Hanjalic}.} \bibinfo{year}{2014}\natexlab{}.
\newblock \showarticletitle{Collaborative filtering beyond the user-item
  matrix: A survey of the state of the art and future challenges}.
\newblock \bibinfo{journal}{\emph{ACM Computing Surveys (CSUR)}}
  \bibinfo{volume}{47}, \bibinfo{number}{1} (\bibinfo{year}{2014}),
  \bibinfo{pages}{1--45}.
\newblock


\bibitem[\protect\citeauthoryear{Victor, De~Cock, and Cornelis}{Victor
  et~al\mbox{.}}{2011}]%
        {victor2011trust}
\bibfield{author}{\bibinfo{person}{Patricia Victor}, \bibinfo{person}{Martine
  De~Cock}, {and} \bibinfo{person}{Chris Cornelis}.}
  \bibinfo{year}{2011}\natexlab{}.
\newblock \showarticletitle{Trust and recommendations}.
\newblock In \bibinfo{booktitle}{\emph{Recommender systems handbook}}.
  \bibinfo{publisher}{Springer}, \bibinfo{pages}{645--675}.
\newblock


\bibitem[\protect\citeauthoryear{West, Waddoups, and Graham}{West
  et~al\mbox{.}}{2007}]%
        {west2007understanding}
\bibfield{author}{\bibinfo{person}{Richard~E West}, \bibinfo{person}{Greg
  Waddoups}, {and} \bibinfo{person}{Charles~R Graham}.}
  \bibinfo{year}{2007}\natexlab{}.
\newblock \showarticletitle{Understanding the experiences of instructors as
  they adopt a course management system}.
\newblock \bibinfo{journal}{\emph{Educational Technology Research and
  Development}} \bibinfo{volume}{55}, \bibinfo{number}{1}
  (\bibinfo{year}{2007}), \bibinfo{pages}{1--26}.
\newblock


\end{thebibliography}

\appendix
\end{document}